\documentstyle[aps,epsfig]{revtex}
\begin{document}

\draft

\title{Photon production in heavy-ion collisions at SPS energies} 

\author{G.Q. Li and G.E. Brown}

\address{Department of Physics, State University of New York at Stony Brook, \\ 
        Stony Brook, NY 11793, USA}

\maketitle

\begin{abstract}

Single photon spectra in heavy-ion collisions at 
SPS energies are studied in the relativistic transport 
model that incorporates self-consistently the change of hadron
masses in dense matter. We separate the total photon
spectrum into `background' arising from the radiative
decays of $\pi^0$ and $\eta$ mesons, and the `themal' one
from other sources. For the latter we include contributions 
from radiative decays of $\rho$, $\omega$, $\eta^\prime$, 
and $a_1$, radiative decays of baryon resonances, 
as well as two-body processes such as 
$\pi\pi\rightarrow \rho\gamma$ and $\pi\rho\rightarrow\pi\gamma$.
It is found that more than 95\% of all photons come from
the decays of $\pi^0$ and $\eta$ mesons, while the 
thermal photons account for less than 5\% of the total photon
yield. The thermal photon spectra in our calculations
with either free or in-medium meson masses do not 
exceed the upper bound set by the experimental 
measurment of the WA80 Collaboration.

pacs: 12.38.Mh, 24.10.Lx, 25.75.-q


\end{abstract}

\section{INTRODUCTION}

One of the primary motivations of ultra-relativistic 
heavy-ion collisons is to create and study in the laboratory
the properties of quark gluon plasma (QGP) as predicted by
quantum chromodynamics (QCD). Experiments have been
carried out at BNL-AGS and CERN-SPS energies, and will be
continued at higher energies of BNL-RHIC and CERN-LHC
colliders \cite{qm96}. The measurement of electromagnetic 
observables such as photon and dilepton spectra 
constitutes a major part of these efforts. The main reason for
this is that the photons and dileptons do not suffer
strong final-state interactions as hadrons do. They can thus
be considered as `penetrating probes' of the initial hot and dense
stages where the QGP might be formed in these ultra-relativistic
heavy-ion collisions \cite{shur78,mc85,kaj86,ruu92}.

So far, the measurement of eletromagnetic observables has 
chiefly been carried out at CERN-SPS energies.
Recent observation of the enhancement of low-mass 
dileptons in central heavy-ion collisions by the
CERES \cite{ceres95a,ceres95b,ceres95c,ceres96a,ceres96b}
and Helios-3 collaborations \cite{helios95} 
has generated a great deal of interest in the heavy-ion 
community. Different dynamical models, such as hydrodynamical 
and transport models, have been used to investigate this phenomenon 
\cite{li95,li96a,li96b,cass95,cass96,gale95,wam95,rapp97,koch96,steele96,steele97,hung96,soll96,red96}. 
Calculations based on `conventional sources' such as Dalitz 
decay and direct vector meson decay that account for 
the dilepton spectra in proton-induced reactions fail
to explain the observed enhancement in heavy-ion collisions.
Various medium effects, such as the dropping vector meson masses
\cite{li95,li96a,li96b,cass95,cass96,hung96,red96} 
as first proposed by Brown and Rho \cite{br91}, the 
modification of rho meson spectral function 
\cite{wam95,rapp97,friman96,weise96}, and the enhanced production of
$\eta$ and/or $\eta^\prime$ \cite{kap96,huang96}, have been proposed
to explain this enhancement. 

Another piece of experimental data from CERN-SPS that have been 
discussed extensively is the single photon spectra in central
S+Au collisions from the WA80 collaboration \cite{wa80}. 
In this experiment, $\pi^0$ and $\eta$ spectra were measured
simultaneously, so that their contributions to single photon
spectra can be subtracted. It was found that the direct photon
excess over these background sources is about 5\% for central
S+Au collisions \cite{wa80a,wa80b,wa80}. 
Similar measurement of inclusive photon spectra has been carried
out by the CERES collaboration \cite{ceres96c}. The results
are very much in agreement with those of WA80, namely, the
inclusive photon spectra can be basically explained by hadronic 
decay, particularly radiative decay of $\pi^0$ and $\eta$ mesons.

Many hydrodynamical calculations 
have been carried out to study photon spectra in these reactions 
\cite{shur94,sinha94,ornik95,dumi95,fai95,cley97,hira97}.
In all these calculations, it is found that the contributions
from the QGP, if it were formed at all in these reactions,
to the single photon yield is negligibly small.
The major contributions to these so-called `thermal' photons,
namely, excessive photons over the background of $\pi^0$ 
and $\eta$ radiative decays, come from hadronic interactions. 

In a number of papers \cite{sinha94,dumi95,hira97}, the absence
of significant thermal photons has been interpreted as an
evidence for the formation of quark gluon plasma. Without
phase transition, the initial temperature of the hadronic gas
was found to be about 400 MeV \cite{sinha94,hira97}. This led to
a large amount of thermal photons from hadronic interaction which
was not observed experimentally. Including the phase transition,
the initial temperature can be lowered to about 200 MeV 
\cite{sinha94,hira97}, because of increased degrees of freedom. 
This lower initial temperature natually reduces the thermal 
photon yield to be in agreement with the WA80 data. There is, 
however, a major caveat in this type of analysis and reasoning, 
namely, the initial temperature in the hadronic scenarios 
depends sensitively on how many hadron resonances are included 
in the analysis. Indeed, in Ref. \cite{soll96}, it was found 
that including a reasonable amount of hadron resonances 
the initial temperature can be reduced to about 250 MeV. 
Furthermore it was shown in Ref. \cite{cley97} that, 
if all the hadron resonances with masses below 2.5 GeV 
are included, the initial temperature can be lowered 
to about 200 MeV. In both papers, it was shown that 
the WA80 photon data can be explained without invoking the
formation of quark gluon plasma. 

There are, up till now, no detailed transport model calculations for
photon spectra in heavy-ion collisions at CERN-SPS energies.
In Ref. \cite{cass97}, results from the Hadron-String dynamics
(HSD), that includes photons from the decay of $\eta^\prime$, 
$\omega$ and $a_1$, were compared with the data and were found
to be far below the upper bound set by the WA80 collaboration.
However, photons can also be produced from the decay of rho mesons,
baryon resonances, and two-body scattering processes such as
$\pi\pi\rightarrow \rho\gamma$ and $\pi\rho\rightarrow \pi\gamma$,
which were neglected in Ref. \cite{cass97}.
Whether the inclusion of these additional sources will
bring the theoretical results close or even beyond the
upper bound of the WA80 data needs to be studied.

One of the purposes of this paper is thus to calculate the
single photon spectra in central S+Au collisions based
on the relativistic transport model and to compare with
hydrodynamical results. The other purpose is to see 
whether the same model that explains the enhancement
of low-mass dileptons, as shown in Refs. \cite{li95,li96a,li96b},
can also explain the lack of signal in the direct photon
measurement. Explaining simultaneously these two correlated
facts constitutes an important test for the consistency 
of the model. 

This paper is arranged as follows: In Section 2 we recall 
briefly the main ingredients of the relativistic transport 
model. In Section 3 we discuss various source for photon 
production, inlcuding meson and baryon decay, as well as two-body
scattering. We shall discuss radiative decay widths
and photon production cross sections in two-body 
processes. The results are discussions are presented
in Section 4. The paper ends with a brief summary
in section 5.
 
\section{THE RELATIVISTIC TRANSPORT MODEL and hadronic observables}

In studying medium effects in heavy-ion collisions, the relativistic 
transport model \cite{ko87,koli96,kkl97} based on the 
Walecka-type model \cite{qhd86} has been quite useful, 
as it provides a thermodynamically consistent 
description of the medium effects through the scalar and vector
fields. In heavy-ion collisions at CERN-SPS energies, 
many hadrons are produced in the initial nucleon-nucleon 
interactions. This is usually modeled by the fragmentation of 
strings, which are the chromoelectric flux-tubes 
excited from the interacting quarks. One successful model 
for taking into account this nonequilibrium dynamics is the RQMD model 
\cite{sorge89}. To extend the relativistic transport model to heavy-ion 
collisions at these energies, we have used as initial conditions the 
hadron abundance and distributions obtained from the string fragmentation 
in RQMD.  

Further interactions and decays of these `primary' hadrons are then 
taken into account as in usual relativistic transport model.
We include non-strange baryons with masses below 1.72 GeV, as well
as $\Lambda$, $\Sigma$, $\Lambda (1405)$ and $\Sigma (1385)$.
For mesons we include $\pi$, $\eta$, $\rho$, $\omega$, $\eta^\prime$,
$a_1$, and $\phi$, as well as $K$ and $K^*(892)$. Baryons are propagated in
their mean fields, which are assumed to be the same for all
non-strange baryons. The mean fields for hyperons are assumed 
to be 2/3 of that for non-strange baryons, based on the simple
quark counting rule. Medium effects on pions and phi mesons 
are neglected, while those on kaons are always included, with 
their mean fields taken to be 1/3 of these for nucleons. 
For other non-strange mesons, we consider two scenarios as in
Refs. \cite{li95,li96a,li96b}, namely, one with medium effects
and one without medium effects. In our model, these 
mesons are treated in the constituent quark model. They thus
feel only the scalar potential that shifts their masses,
since the vector potential on the (constiutent) quark is canceled by
that on the (constituent) antiquark.

In addition to propagation in mean fields, hadrons also 
under stochastic two-body collisions.
For baryon-baryon interactions, we include both elastic
and inelastic scattering for nucleons, $\Delta (1232)$,
$N(1440)$ and $N(1535)$. Their cross sections are either
taken from Refs. \cite{arndt82,wolf93} or obtained using
the detailed balance procedure \cite{bert91}. The meson-baryon
interactions are modeled by baryon resonace formation and decay.
For example, the interaction of a pion with a nucleon
proceeds through the formation and decay of $\Delta (1232)$,
$N(1440)$, ..., N(1720). The formation cross sections
are taken to be of the relativistic Breit-Wigner form.
The meson-meson interactions are either formulated
by the resonance formation and decay when the intermediate 
meson is explicitly included in our model or treated as
a direct elastic scattering with a cross section estimated
from various theoretical models. Of particular importance
for photon production are the pion-pion and pion-rho collisions,
which are dominated by rho and $a_1$ meson formation,
respectively. 

As mentioned in the Introduction, the observed enhancement of low-mass
dileptons in heavy-ion collisions can be explained by 
including dropping vector meson masses. It is useful to see
whether the inclusion of the dropping vector meson mass
is also consistent with the real photon data from the WA80
collaboration. For this purpose we will also calculate
photon spectra in the scenario of dropping meson masses.
Following Ref. \cite{li95,li96a,li96b}, we extend the 
Walecka model from the coupling of nucleons to scalar 
and vector fields to the coupling of light quarks to these 
fields, using the ideas of the meson-quark coupling model 
\cite{thomas94}. For a system of nucleons, pseudoscalar mesons,
vector mesons, and axial-vector mesons at temperature 
$T$ and baryon density $\rho _B$, the scalar field 
$\langle\sigma\rangle$ is determined self-consistently from 
\begin{eqnarray}
m_\sigma^2\langle \sigma\rangle &=&{4g_\sigma\over (2\pi )^3}\int d{\bf k} 
{m_N^*\over E^*_N}\Big[{1\over \exp ((E^*_N-\mu _B)/T)+1}
+{1\over \exp ((E^*_N+\mu _B)/T)+1}\Big]\nonumber\\
&+&{0.45g_\sigma\over (2\pi )^3}\int d{\bf k} {m_\eta^*\over E_\eta ^*}
{1\over \exp (E_\eta ^*/T)-1}+
{6g_\sigma\over (2\pi )^3}\int d{\bf k} {m_\rho^*\over E_\rho ^*}
{1\over \exp (E_\rho ^*/T)-1}\nonumber\\
&+&{2g_\sigma\over (2\pi )^3}\int d{\bf k} {m_\omega^*\over E_\omega ^*}
{1\over \exp (E_\omega ^*/T)-1}
+{6\sqrt 2 g_\sigma\over (2\pi )^3}\int d{\bf k} {m_{a_1}^*\over E_{a_1}^*}
{1\over \exp (E_{a_1}^*/T)-1},
\end{eqnarray}
where we have used the constituent quark model relations for 
the nucleon and vector meson masses \cite{thomas94}, 
i.e., $m_N^*=m_N-g_\sigma\langle \sigma\rangle ,
~m_{\rho ,\omega}^*\approx m_{\rho ,\omega}
-(2/3)g_\sigma\langle\sigma\rangle$, the quark 
structure of the $\eta$ meson in free space which leads to 
$m_\eta^*\approx m_\eta -0.45g_\sigma\langle\sigma\rangle$, and 
the Weinberg sum rule relation between the rho-meson and $a_1$ 
meson masses, 
i.e, $m_{a_1}^*\approx m_{a_1}-(2\sqrt 2/3)g_\sigma\langle\sigma\rangle$.
Recently we found that the use of a refined model, the effective
chiral Lagrangian of \cite{fst},
leads to essentially the same results for dilepton spectra \cite{li97}.

This model describes the hadron observables quite well. In Fig. 1
we show the pseudorapidity distribtion of charged particles
(upper window) and transverse momentum spectra of pions
(lower window) for central S+Au collisions. They are
found to be in good agreement with the experimental
data from the WA80 \cite{wa80z,wa80n} and the
CERES collaboration \cite{ceresn}. Comparisons with
other hadron observables, such as proton rapidity distribution
and transverse mass spectra can be found in Ref. \cite{li96b}.
  
\section{PHOTON PRODUCTION: radiative widths and cross sections}

The majority of single photons come from the decay of $\pi^0$ and
$\eta$ mesons at freeze out. The background photon spectra from
their decay can thus be evaluated at the end of the 
transport model simulation by using the following 
branching ratio \cite{pdata}
\begin{eqnarray}
B_{\pi^0\rightarrow \gamma\gamma}
={\Gamma _{\pi^0\rightarrow \gamma\gamma}\over
\Gamma _\pi^0}=0.988, ~~
B_{\eta\rightarrow \gamma\gamma}
={\Gamma _{\eta\rightarrow \gamma\gamma}\over
\Gamma _\eta}=0.3925.
\end{eqnarray}

For the `thermal' photon spectra we include the decay of $\rho$, $\omega$,
$\eta^\prime$, and $a_1$ mesons, the decay of all the baryon 
resonances in our model, as well as two-body processes such
as $\pi\pi\rightarrow \rho \gamma$ and $\pi\rho\rightarrow \pi\gamma$.
The decay widths for $\rho^0\rightarrow \pi^+\pi^-\gamma$ and
$\rho^\pm\rightarrow \pi^\pm\pi^0\gamma$ are taken from
\cite{sig63}, which explains the measured width for $\rho^0\rightarrow
\pi^+\pi^-\gamma$. The mass dependence of these widths are shown in Fig. 2. 
The open circle gives the experimental data for
$\rho^0\rightarrow \pi^+\pi^-\gamma$ \cite{pdata}. 
Note that in transport models, rho mesons can have different 
masses because of its broad mass distribution, so mass dependent
widths are used.

The omega meson radiative decay width $\Gamma _{\omega\rightarrow
\pi^0\gamma}$ was studied in Ref. \cite{kay84} based
on the chiral Lagrangian that includes the Wess-Zumino anomalous
term. The width was found to be proportional to the
pion momentum in the $\omega$ rest frame, namely,
\begin{eqnarray}
\Gamma _{\omega\rightarrow \pi^0\gamma} (M)
=C |{\bf p}_\pi |^3 \approx 0.01316 |{\bf p}_\pi |^3,
\end{eqnarray}
where $|{\bf p}_\pi | = (M^2-m_\pi^2) /(2M) $ and the coefficient 
$C$ is determined from the measured width \cite{pdata}.

There exists only one experimental measurement of 
the $a_1$ radiative decay width, using the reverse
process of $a_1$ production in pion-nucleus collisions \cite{ziel84}.
The width is found to be $\Gamma _{a_1\rightarrow \pi \gamma} 
=0.64\pm 0.246$, which was used in Ref. \cite{cass97} in
calculating $a_1$ contribution to photon production.
In Ref. \cite{shur92}, an effective Lagrangian was proposed
to calculate this width, which turns out to be about 1.4 MeV
using the vector dominance assumption. This is close to
the prediction of non-relativistic quark model of about 1.0-1.6
MeV \cite{ros81}. In view of large uncertainties, we use
$\Gamma _{a_1\rightarrow \pi \gamma} (m_{a_1}) = 1$ MeV as 
in Refs. \cite{li96a,li96b}. According to Ref. \cite{shur92},
in the limit that pion momentum is much greater than its mass,
the $a_1$ radiative decay width is given by
\begin{eqnarray}
\Gamma _{a_1\rightarrow \pi\gamma} (M)
=C |{\bf p}_\pi |^5 \approx 0.01214 |{\bf p}_\pi |^5,
\end{eqnarray}
where the coefficient $C$ is determined by using 
$\Gamma _{a_1\rightarrow \pi \gamma} (m_{a_1}) = 1$ MeV.

Photons can also be produced from the decay of baryon
resonaces. These contributions are usually neglected
in hydrodynamical calculations \cite{sinha94,hira97}.
The radiative decay widths for baryon resonances included
in our transport model have been measured experimentally
through photon-nucleon and/or pion-nucleon interactions 
\cite{pdata}. We use these widths in our calculation of their
contributions to single photon spectra. In addition, $\Sigma^0$
decay dominantly into $\Lambda\gamma$. But since the mass
difference between $\Sigma^0$ and $\Lambda$ is not very
large, photons from $\Sigma^0$ decay usually have low transverse
momenta, as will be shown below.

For contributions from two-body processes, we include
$\pi\pi\rightarrow \rho\gamma$ and $\pi\rho\rightarrow \pi\gamma$,
which are the most important ones in the temperature
region relevant for SPS energies \cite{kapu91,song93}. 
These cross sections were first evaluated in Ref. \cite{kapu91}, 
with an effective Lagragian including $\pi$, $\eta$, $\rho$, 
and $\omega$ mesons, which will be used in the present work. 
The cross section for $\pi\pi\rightarrow \rho\gamma$ 
is shown in Fig. 4, for three different rho meson masses.
As the rho meson mass is reduced, the threshold and also the
magnitude of the cross section are reduced. The cross section
for $\pi\rho\rightarrow \pi\gamma$ is shown in Fig. 5 for
three different rho meson mass. When the rho meson mass is reduced,
the available center-of-mass energy is reduced and the cross
section increases.

The importance of the $a_1$ meson was 
investigated in Refs. \cite{shur92,song93}. Since in our
transport model the processes $\pi\rho\rightarrow a_1$ and
$a_1\rightarrow \pi\gamma$ are explictly included, we need
not to include the effects of $a_1$ on the direct 
$\pi\rho\rightarrow \pi\gamma$ process. Otherwise there will
be double counting. The effects of the $a_1$ on
$\pi\pi\rightarrow \rho\gamma$ were found to be appreciable
for photons with energies greater than about 0.5 GeV \cite{song93}.
The contributions of $\pi\pi\rightarrow \rho\gamma$ to these
`hard' photons are, however, extremely small (about one to two orders
of magnitude smaller than those from  $\pi\rho\rightarrow \pi\gamma$).
This justifies our use of the cross sections given in Ref. \cite{kapu91}
that did not include the $a_1$ meson in the model.

\section{PHOTON PRODUCTION: results and discussions}

We present in this section the single photon spectra
in central S+Au collisions at 200 AGeV. The
acceptance of the WA80 collaboration is taken into
account, namely, we include photons with pseudo-rapidity 
of $2.1<\eta<2.9$. In Fig. 6 we show the
background  photon spectra from the decay of
$\pi^0$ (dotted line) and $\eta$ meson (dashed line).
It is seen that contribution from $\pi^0$ decay
far dominates the background.

The contributions to the so-called `thermal' photons
from the decay of mesons and baryons, and from two-body
scattering are shown in Fig. 7. The $\omega$ radiative
decay is found to be the most important source for photons
with transverse momenta above about 0.5 GeV. This
is in agreement with the finding of Ref. \cite{cass97},
although our results are somewhat larger than those
of Ref. \cite{cass97}. This difference might be traced
back to the difference in the `primary' $\omega$ 
meson abundances in the RQMD and the HSD models.
The contribution from the $a_1$ radiative decay is
comparable to that from direct $\pi\rho\rightarrow \pi\gamma$.
This is in agreement with the conclusion of Ref. \cite{song93}
that including $a_1$ the $\pi\rho$ contribution increases 
by about a factor 2-3 in the relevant temperature and photon
energy region.

The contributions from $\eta^\prime$ and $\Sigma^0$
radiative decays are restricted to photons with transverse 
momenta below 0.5 GeV, because the mass differences between hadrons
in the intial and final states in these processes are small.
Photons with transverse momenta below 0.2 GeV come chiefly
from the decay of $\Sigma ^0$ and $\pi\pi$ scattering.
The reaction $\pi\pi\rightarrow \rho\gamma$ is endothermic,
with most of the available energy going into the rho meson mass.
This cross section actually diverges as the photon energy
goes to zero. We have included a low-energy cut-off as
in Ref. \cite{kapu91}. The choice of this cut-off parameter
affects basically only the photons with transverse momenta
below about 0.1 GeV, where no experimental data are available.

In Fig. 8 we compare our results with the upper bound of the
WA80 collaboration. More than 80\% of thermal photons come
from meson decay, of which more than half come from the $\omega$
radiative decay. Overall, our results are well below
the WA80 upper bound for photons with transverse momenta below
1 GeV. For higher tranverse momenta, our results touch the
upper bound of the experimental data. Our results are actually 
very close to those of Ref. \cite{soll96} with the parameter set 
named EOS H, that includes about the same number of hadron 
degrees of freedom as in our model. The conclusion is thus that
the WA80 single photon data do not necessary imply
the formation quark gluon plasma. Of course,
the possibility of quark gluon plasma
formation is not ruled out.
 
Next we discuss the results obtained with in-medium meson
masses. The separate contributions from the decay
of mesons and baryons, and two-body processes are
shown in Fig. 9. There is not much change in the
meson and baryon decay contributions. With dropping
meson masses, the abundances of $\rho$ and $a_1$ mesons
increase \cite{li96a,li96b}. This is, however, compensated
by the decreases in their radiative decay widths (see Figs.
2 and 3). The comparison of total thermal photon
spectra with the experimental data is
given in Fig. 10. In the case of dropping meson 
masses, our results do not exceed the upper bound
of the WA80 data. 

Finally, in Fig. 11 we show the ratio of total 
(sum of thermal and background) photon to the
background photon. In both and free meson mass 
and in-medium meson mass cases, the thermal photon
accounts for less than 5\% of all the single
photons, in agreement with the experimental
observation, and the simple estimation by Tserruya 
in Ref. \cite{ceres95c}. An important difference 
between dilepton and real photon measurements is
that in the former case, the contamination from
$\pi^0\rightarrow \gamma e^+e^-$ can, in principle,
be removed since these dileptons are restricted to masses
below $m_{\pi^0}$, while in the real photon case,
the $\pi^0\rightarrow \gamma\gamma$ contributes to
photons with any momentum. The subtraction of
the background, and hence the measurement of 
interesting signals, are much more subtle in the
real photon case \cite{ceres95c}.

\section{Photon production: comparison with other calculations}

The WA80 single photon data have been looked at by several
groups, mostly based on hydrodynamical models 
\cite{steele97,hung96,soll96,shur94,sinha94,ornik95,dumi95,fai95,hira97,cass97}
Here we compare our results that are based on the relativistic
transport model with some of these results. 
In Fig. 12, we show our results together with those
from \cite{soll96} and \cite{sinha94}. In Ref. \cite{soll96}
different equations of state with and without quark gluon
plasma formation were considered. Shown in the figure by dotted line
is their results based on EOS H. This equation of state
included about the same number of hadron degrees of freedom as
in our transport model. It is very interesting to see that
their results are very similar to ours, although the dynamical
models used are quite different. In Ref. \cite{sinha94}
two different equations of state, one with quark gluon
plasma formation and one with pure pionic gas, were
considered. Shown in the figure is their results 
based on the pionic gas equation of state. Because of
very limited number of hadronic degrees of freedom in
this equation of state, they needed a very high 
initial temperature to account for the final observed
hadron abundances. This led to large photon yield.

In Refs. \cite{steele96,steele97}. Steele {\it et al.}
studied photon as well as dilepton production at SPS
energy based on the master equation formalism.
They have considered two scenarios, one with pions
only and the other includes nucleon as well. 
In order to make a fair comparison, we
show in the upper window of Fig. 13 the photon
spectra obtained in our calculation with the
decays of $\rho$ and $a_1$ and two-body processes
of $\pi\pi\rightarrow \rho\gamma$ and $\pi\rho\rightarrow
\pi\gamma$, and the results of Ref. \cite{steele97} 
excluding nucleons. We can see that their results
are very similar to ours. In the lower part
of Fig. 13 we compare our results including baryon resonance
decay with those of Ref. \cite{steele97} that included nucleon 
effects. The results are again quite consistent with ours.
Note that the results of Ref. \cite{steele97} is very
sensitive to the initial nucleon density. The results shown
in the lower part of Fig. 13 were obtained with an initial
nucleon density $\rho_N=0.7 \rho_0$, which is quite
appropriate for S+Au collisions at SPS energies.
 
\section{SUMMARY}
 
In summary, we studied single photon spectra in central 
S+Au collisions at SPS energies using the relativistic 
transport model that has been used to study dilepton spectra 
in the same reactions.  We included photons from the background
sources of $\pi^0$ and $\eta$ decays, as well as thermal sources
such as meson decays, decays of baryon resonances, and 
two-body processes. We found that more than 95\% of single
photons come from the decays of $\pi^0$ and $\eta$. The 
thermal photons account for only less than 5\% of all single
photons, in agreement with the experimental observation made by
the WA80 and CERES collaborations. We compared our thermal photon 
spectra with the experimental upper bound extracted by
the WA80 collaboration. It is seen that in both the free
meson mass and in-medium meson mass cases, our results
do not exceed the experimental upper bound. 
This indicates that our model can explain both the
enhancement of low-mass dileptons and the lack
of signal in the real photon measurement.
Our results are also in agreement with the conclusions 
of hydrodynamical calculations of Refs. \cite{soll96,cley97} 
that included a sufficient number of hadron degrees of freedom. 
Therefore the WA80 single photon data do not necessary imply
the phase transition to the quark gluon plasma, as
claimed in Refs. \cite{sinha94,dumi95,hira97} based on
a limited number of hadron degrees of freedom.

Finally, we note that the elementary radiative decay widths and
photon prodcution cross sections used in this study have been
taken from different models. In principle, most of them
can be evaluated consistently within the same model, such
as the hidden gauge theory of \cite{hidd}. In this model,
one can also study the effects of the change of gauge coupling 
constant on photon production in hot and dense matter. 
Work in this direction is in progress and will be reported 
elsewhere \cite{adam97}.

\vskip 1cm

We thank M.A. Halasz, C.M. Ko, J.V. Steele and I. Zahed
for discussions, and I. Tserruya for a careful reading of
the manuscript and useful suggestions. This work is supported 
in part by the Department of Energy under grant No. De-FG-88ER40388.

\newpage

\centerline{\bf Figure Captions}

Fig. 1: Pseudorapity distribution of charged hadrons (upper 
window) and pion transverse momentum spectra (lower window)
in central S+Au collisions at 200 AGeV.

Fig. 2: Decay width for $\rho\rightarrow \pi\pi\gamma$ as 
a function of rho meson mass. The open circle is experimental
data for $\rho^0$.

Fig. 3: Radiative decay widths of $\omega$ (upper window)
and $a_1$ (lower window) mesons as a function of their masses.
The open circles are experimental data.

Fig. 4: Cross sections for $\pi\pi\rightarrow \rho\gamma$
from Ref. \cite{kapu91}, for three different masses.

Fig. 5: Cross sections for $\pi\rho\rightarrow \pi\gamma$
from Ref. \cite{kapu91}, for three different
masses.

Fig. 6: Backgroud single photon spectra from $\pi^0$ and $\eta$
decays in central S+Au collisions.

Fig. 7: thermal single photon spectra from $\rho$, $\omega$,
$\eta^\prime$ and $a_1$ decays (left window),
from $N^*$, $\Delta^*$ and $\Sigma ^0$ decays (middle window), 
and from $\pi\pi$ and $\pi\rho$ scattering (right
window) in central S+Au collisions.

Fig. 8: thermal single photon spectra meson decay (dotted line),
baryon decay (short-dashed), and two-body scattering (long-dashed)
in central S+Au collisions. The solid circles give the upper
bound from the WA80 collaboration \cite{wa80}.

Fig. 9: Same as Fig 7, the results with in-medium meson masses.

Fig. 10: Same as Fig. 8, the results with in-medium meson masses.

Fig. 11: The ratio of total photon spectra to the backgroud
photon spectra in central S+Au collisions. The solid and dashed
histograms are the results with in-medium and free meson
masses, respectively. The solid circles are the experimental
data from the WA80 collaboration \cite{wa80}.

Fig. 12: Comparison of our results with those of
Ref. \cite{soll96} (dotted line) and
Ref. \cite{sinha94} (long-dashed line).

Fig. 13: Comparison of our results with those of
Ref. \cite{steele97}. Upper window: pions only,
lower window: pions and baryons.

\newpage
\begin{figure}
\begin{center}
\epsfig{file=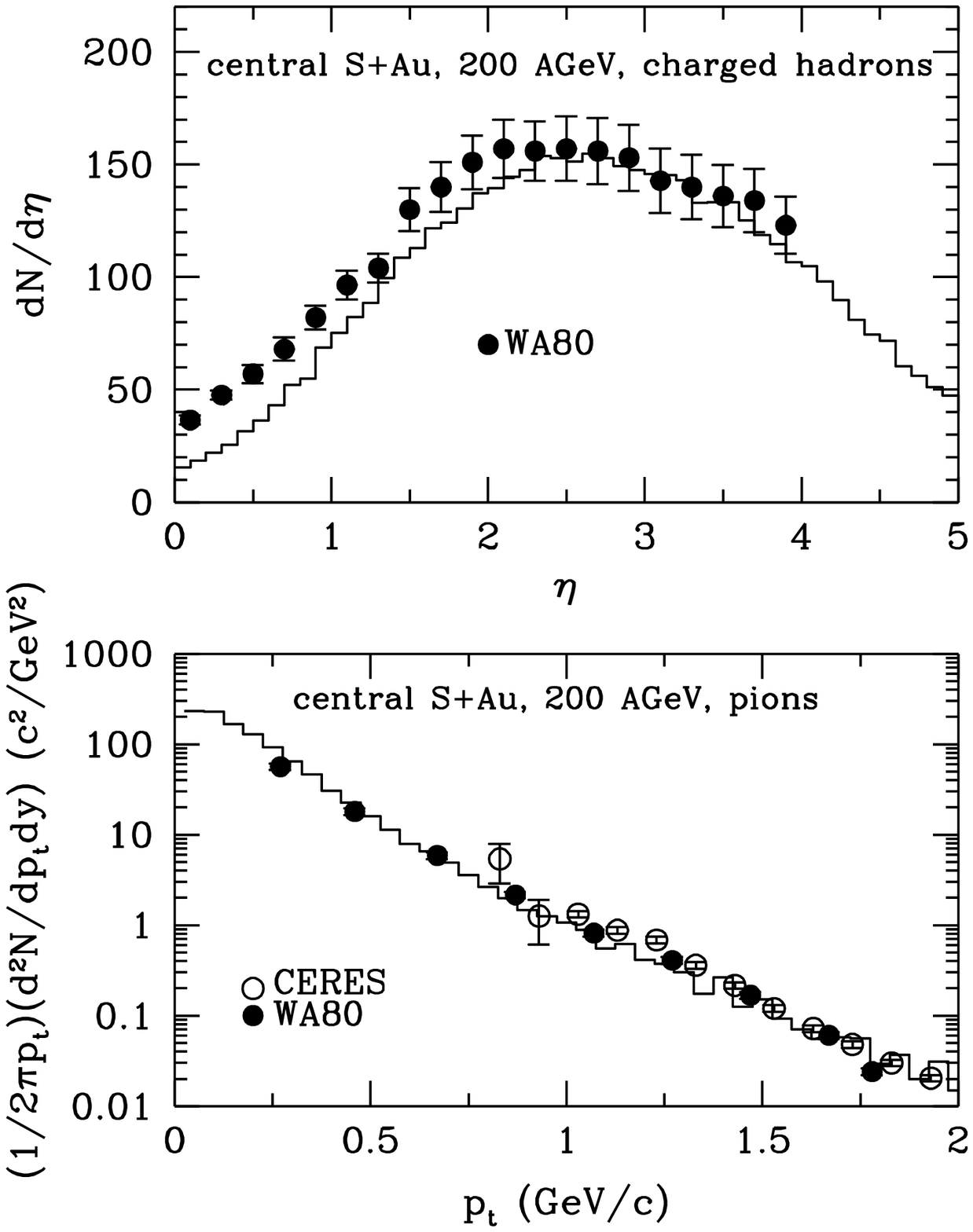,height=6in,width=6in}
\end{center}
\end{figure}
 
\newpage
\begin{figure}
\begin{center}
\epsfig{file=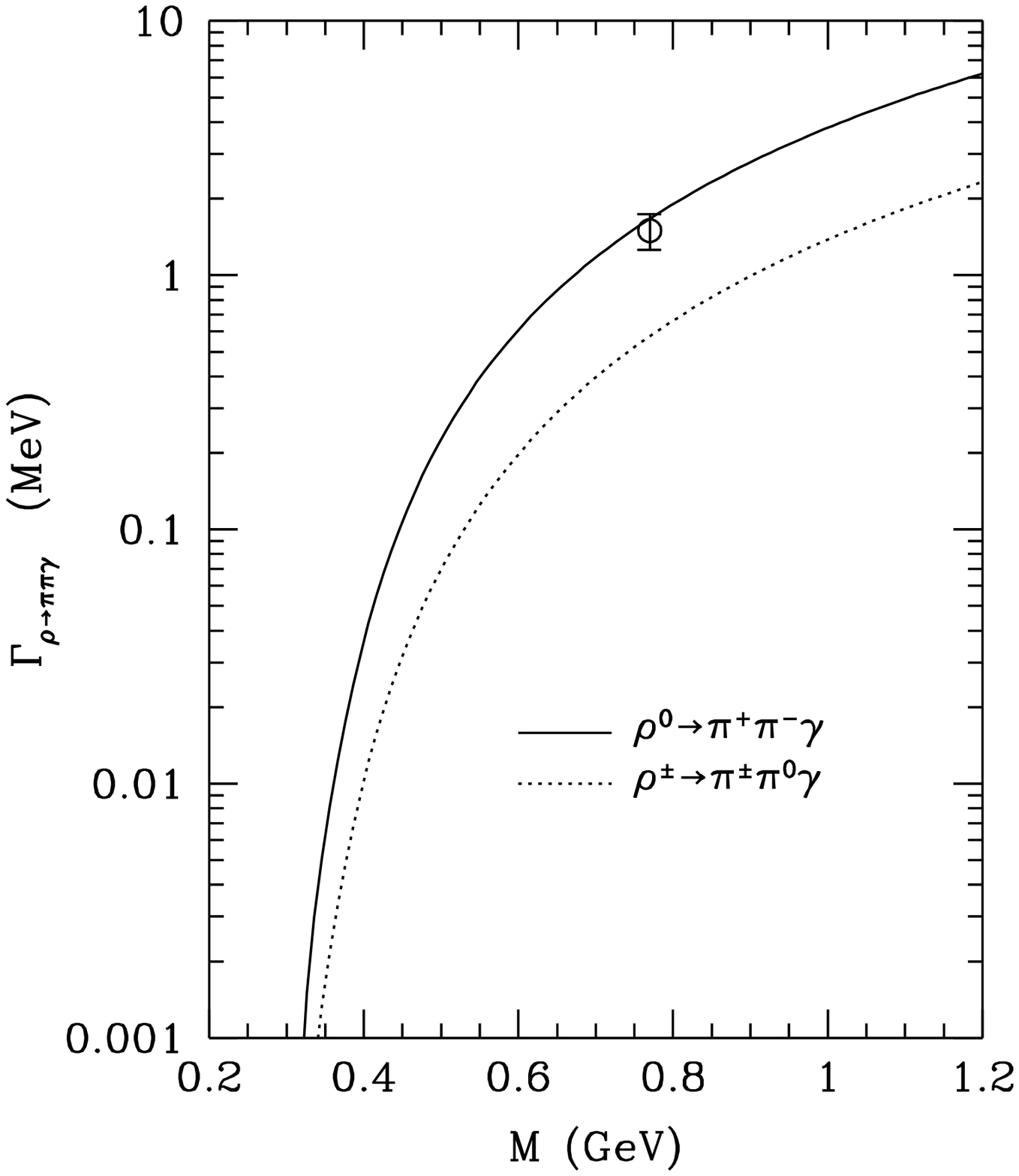,height=6in,width=6in}
\end{center}
\end{figure}
 
\newpage
\begin{figure}
\begin{center}
\epsfig{file=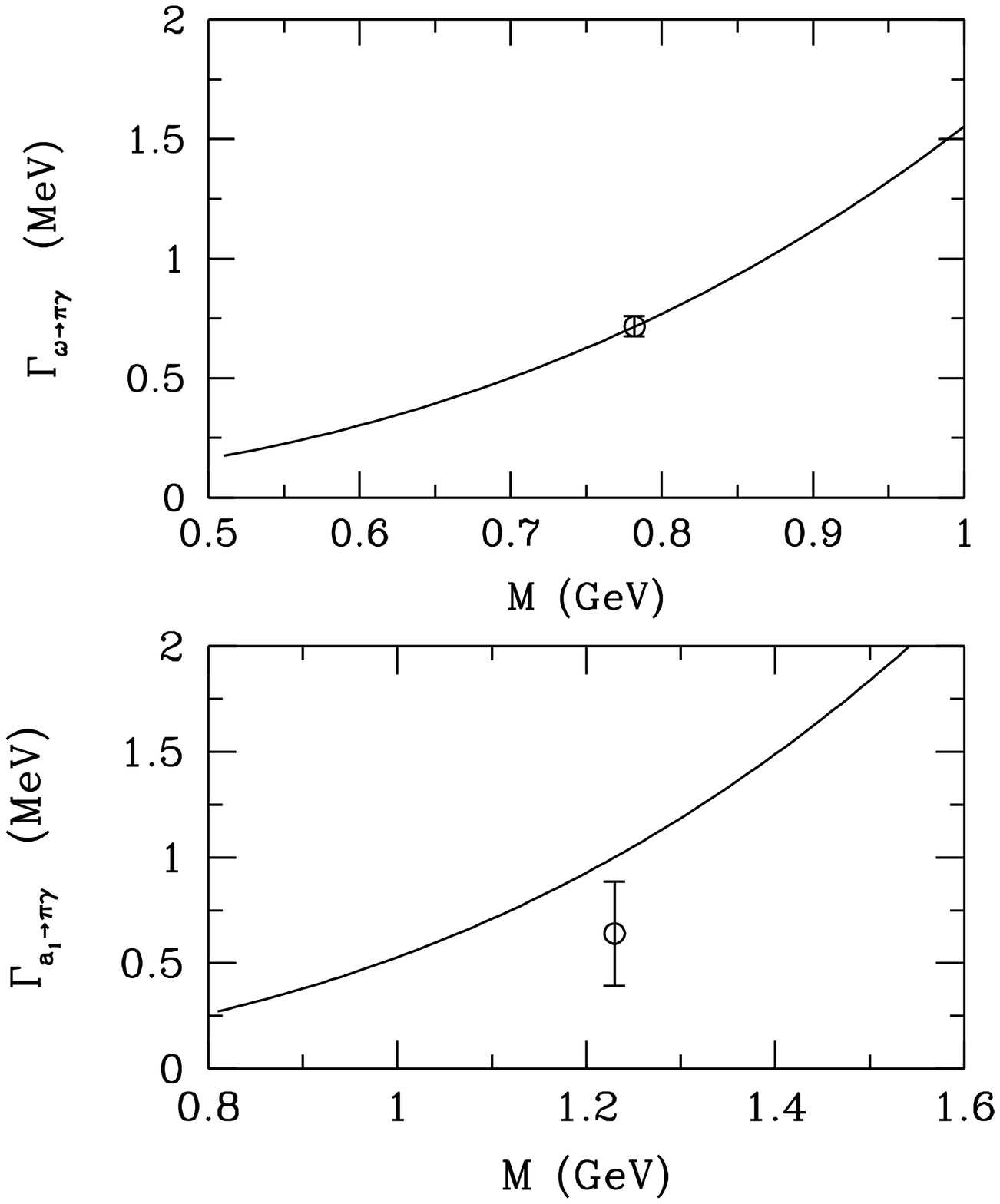,height=6in,width=6in}
\end{center}
\end{figure}
 
\newpage
\begin{figure}
\begin{center}
\epsfig{file=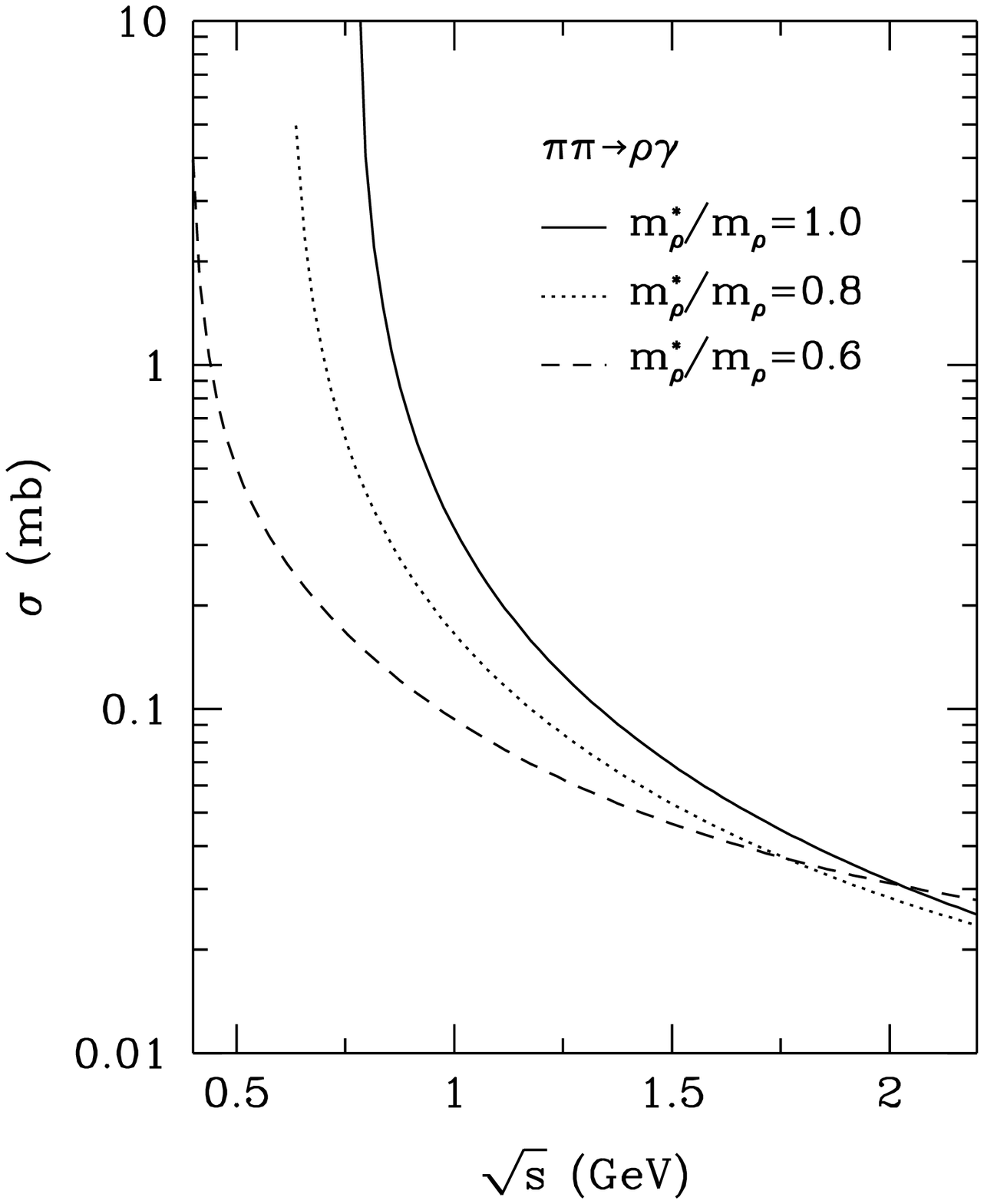,height=6in,width=6in}
\end{center}
\end{figure}
 
\newpage
\begin{figure}
\begin{center}
\epsfig{file=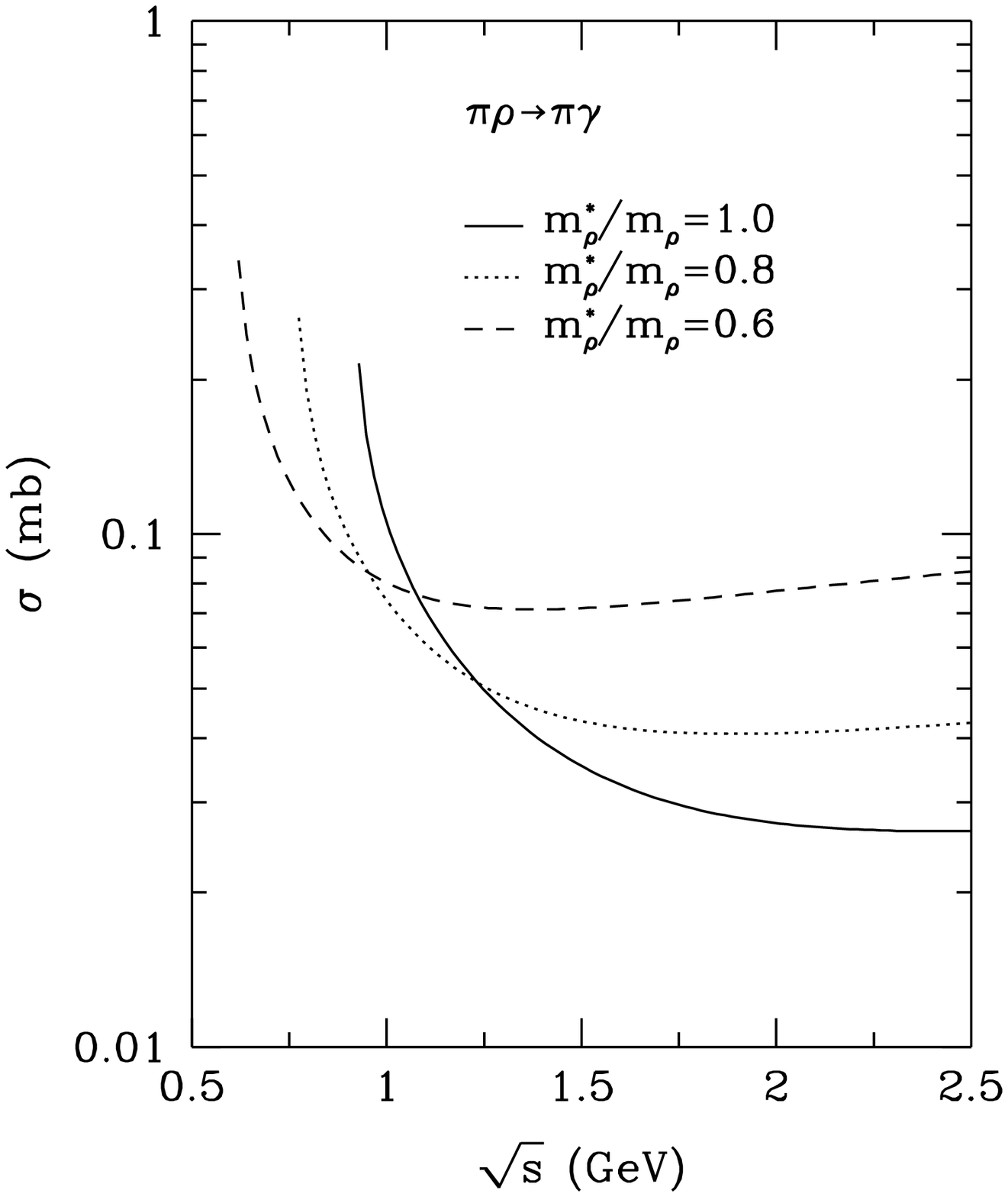,height=6in,width=6in}
\end{center}
\end{figure}
 
\newpage
\begin{figure}
\begin{center}
\epsfig{file=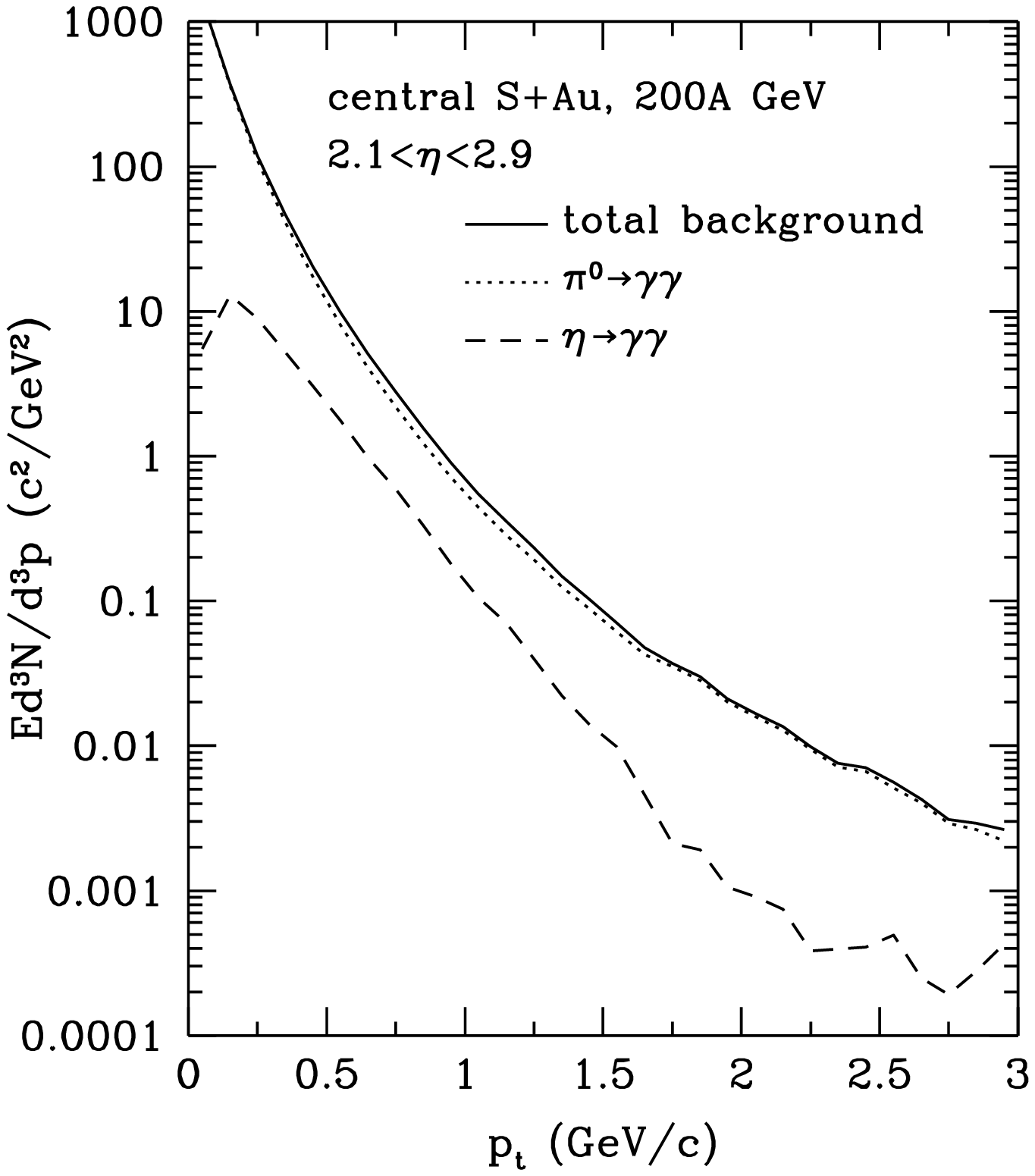,height=6in,width=6in}
\end{center}
\end{figure}
 
\newpage
\begin{figure}
\begin{center}
\epsfig{file=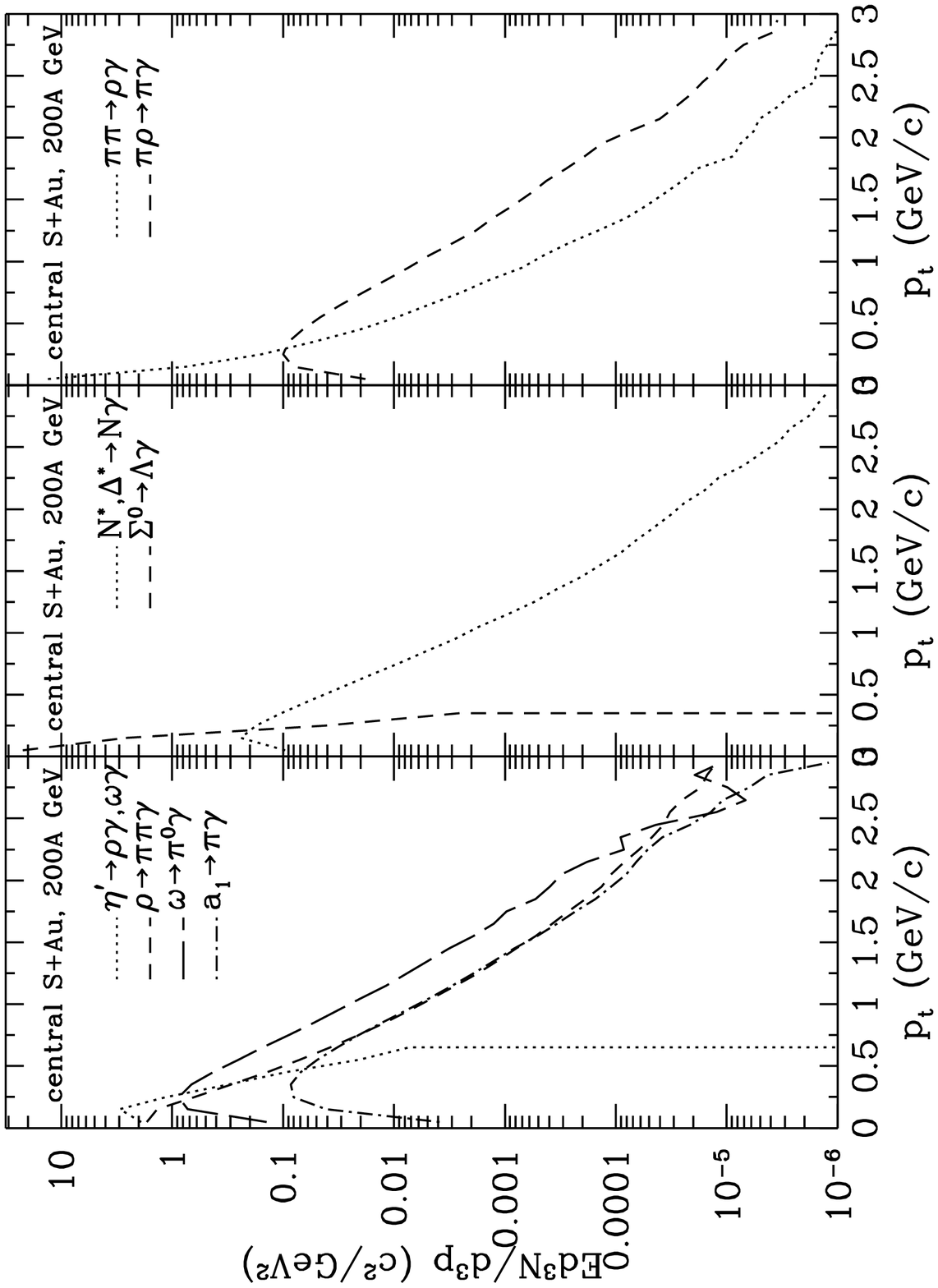,height=6in,width=6in}
\end{center}
\end{figure}
 
\newpage
\begin{figure}
\begin{center}
\epsfig{file=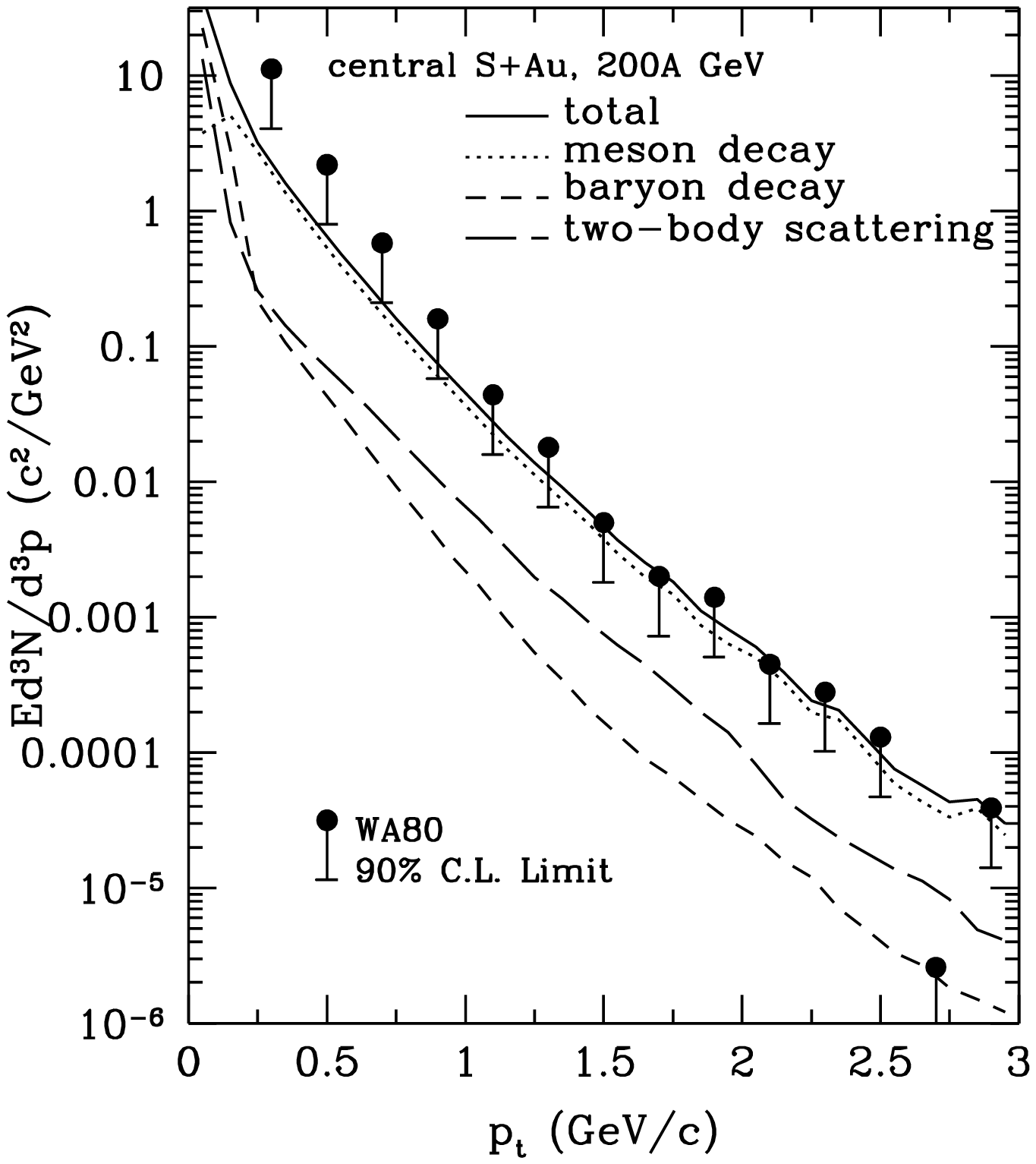,height=6in,width=6in}
\end{center}
\end{figure}
 
\newpage
\begin{figure}
\begin{center}
\epsfig{file=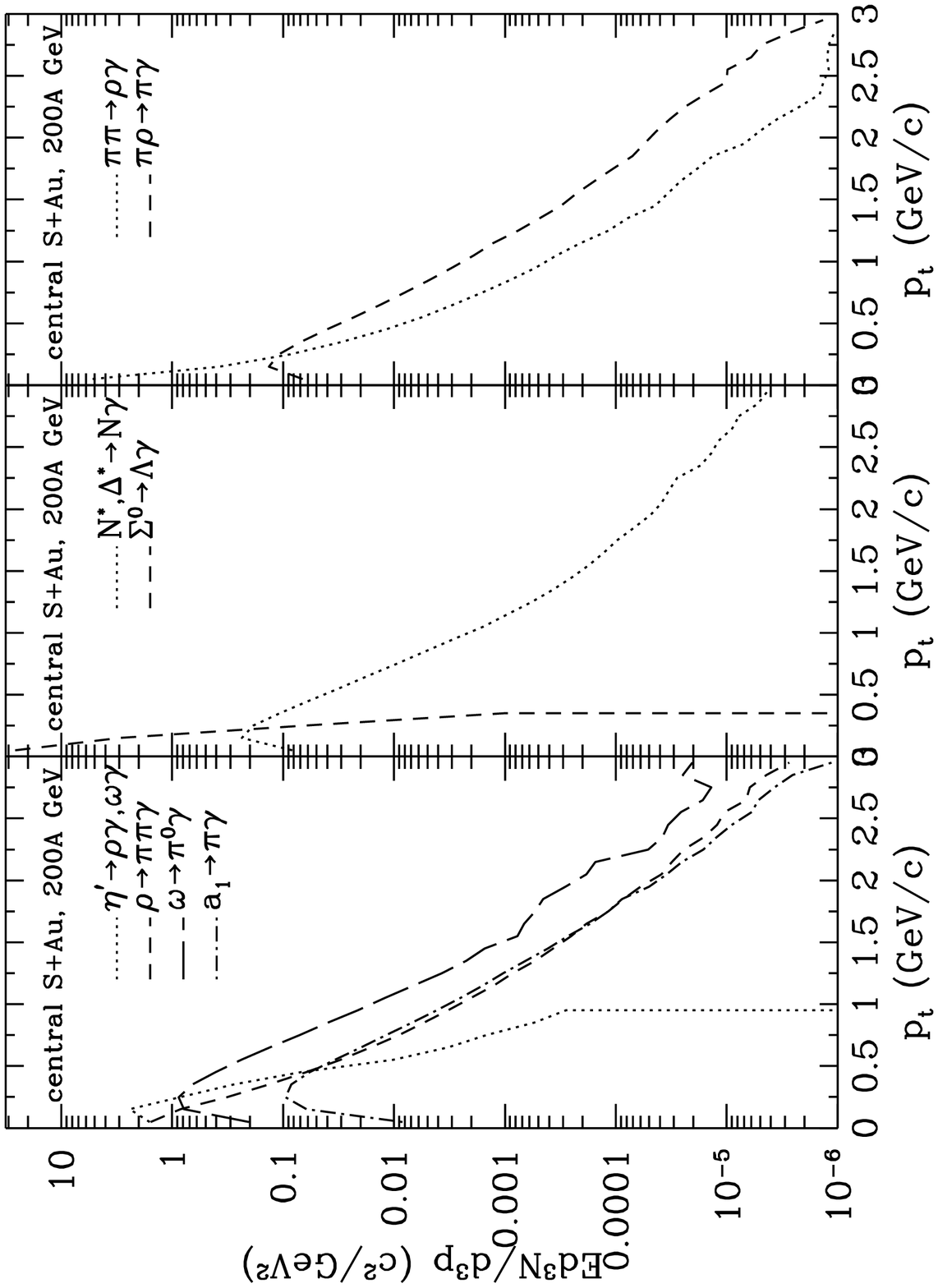,height=6in,width=6in}
\end{center}
\end{figure}

\newpage
\begin{figure}
\begin{center}
\epsfig{file=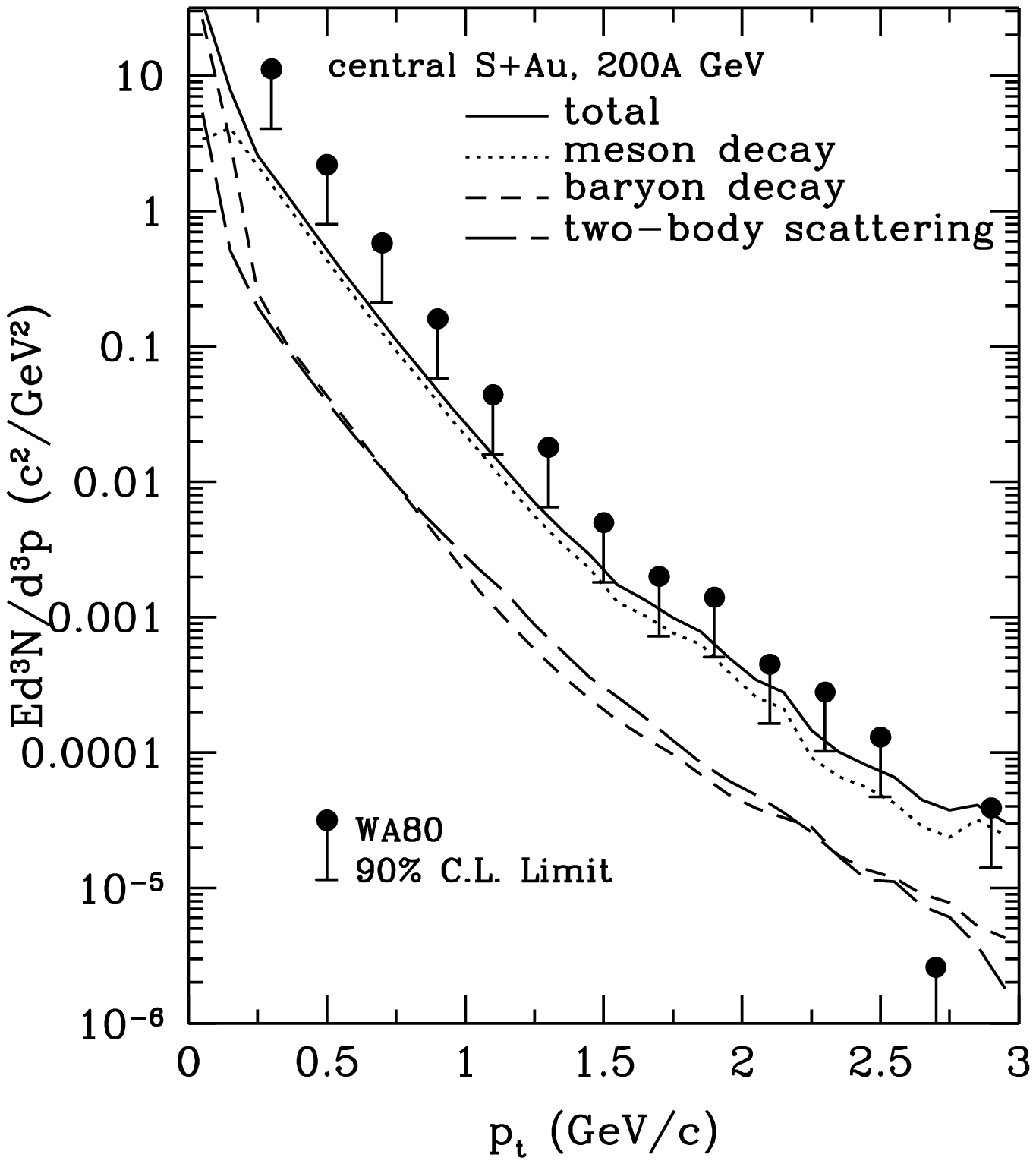,height=6in,width=6in}
\end{center}
\end{figure}

\newpage
\begin{figure}
\begin{center}
\epsfig{file=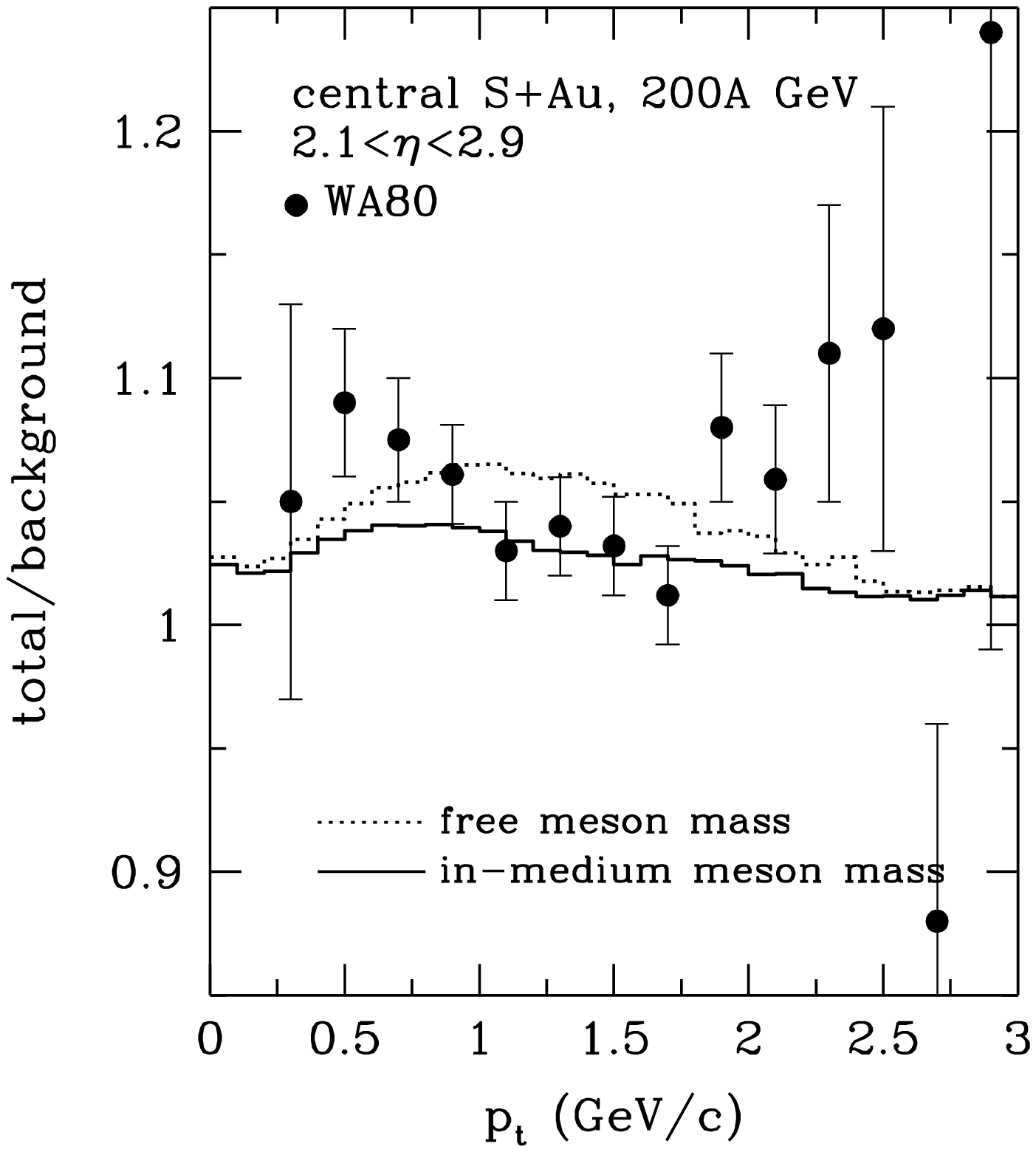,height=6in,width=6in}
\end{center}
\end{figure}

\newpage
\begin{figure}
\begin{center}
\epsfig{file=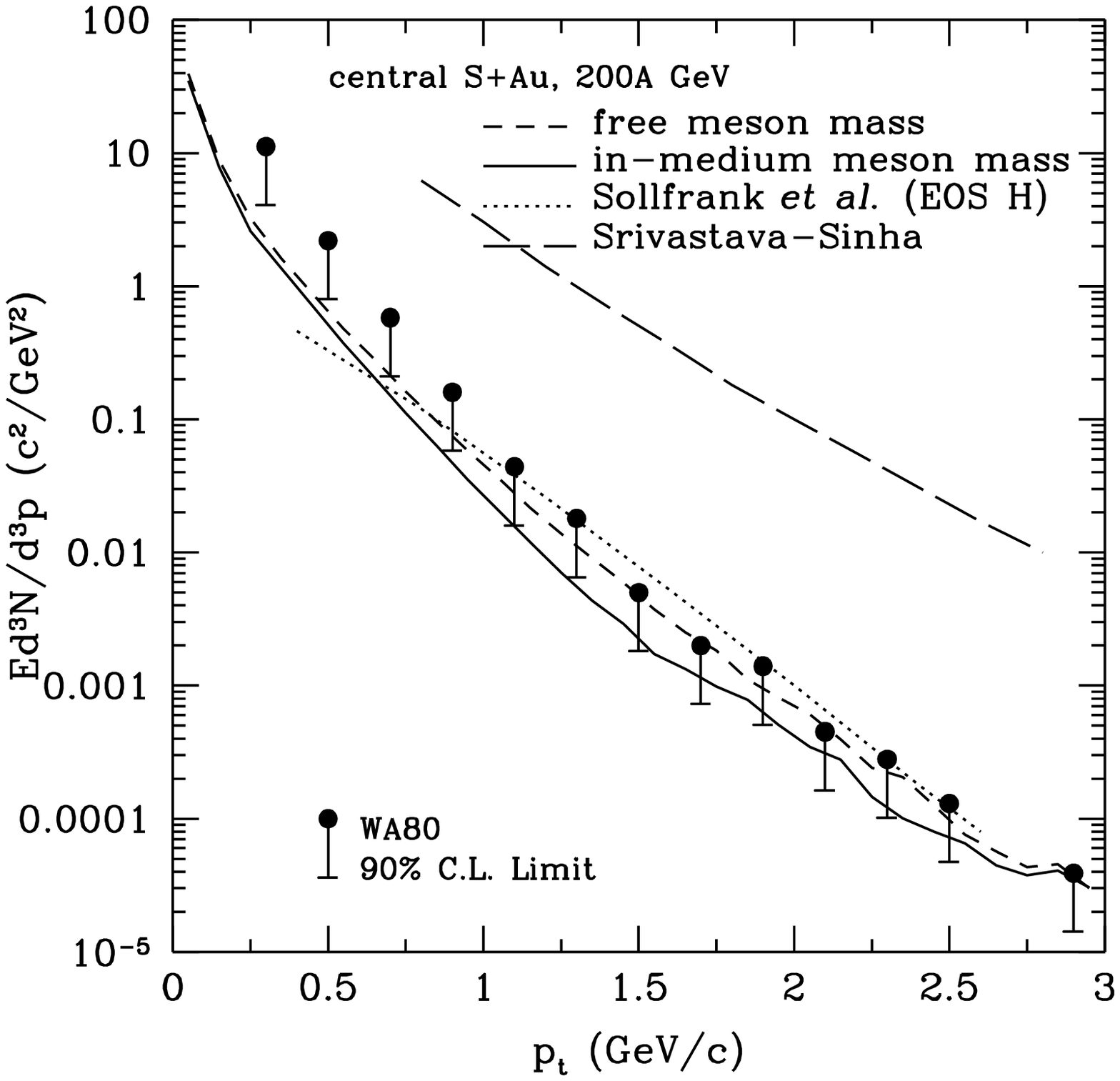,height=6in,width=6in}
\end{center}
\end{figure}

\newpage
\begin{figure}
\begin{center}
\epsfig{file=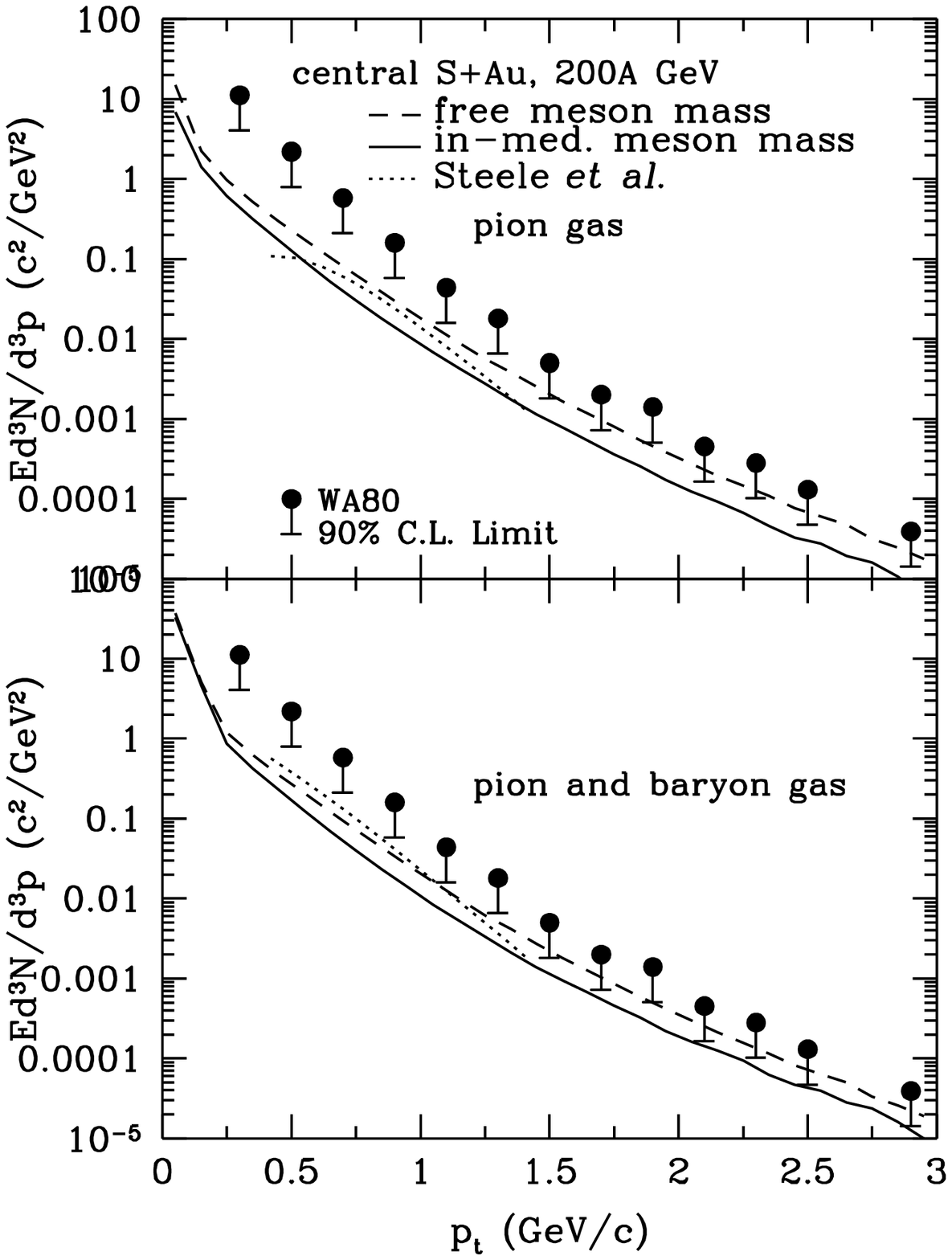,height=6in,width=6in}
\end{center}
\end{figure}

\end{document}